\documentclass[journal,a4paper]{IEEEtran}
\usepackage{multirow}
\usepackage{graphicx}
\usepackage{amsmath}
\usepackage[psamsfonts]{amssymb}
\usepackage{amsxtra}
\usepackage{threeparttable}

\begin{document}

\title{Fast and Accurate Frequency Estimation Using Sliding DFT}
\author{{Anit Kumar Sahu$^1$, Mrityunjoy
Chakraborty$^2$}\\
Department of Electronics and Electrical Communication
Engineering\\
Indian Institute of Technology, Kharagpur, INDIA\\
E.Mail : $^1\;$anit.sahu@gmail.com,
$^2\;$mrityun@ece.iitkgp.ernet.in}

\maketitle
\thispagestyle{empty}

\begin{abstract}
Frequency Estimation of a complex exponential is a problem 
relevant to a large number of fields. In this paper a 
computationally efficient and accurate frequency estimator 
is presented using the guaranteed stable Sliding DFT which 
gives stability as well as computational efficiency. The estimator 
approaches Jacobsen's estimator and Candan's estimator for 
large N with an extra correction term multiplied to it for 
the stabilization of the sliding DFT. Simulation results 
show that the performance of the proposed estimator were 
found to be better than Jacobsen's estimator 
and Candan's estimator.\end{abstract}

\section{Introduction}
The frequency estimation of a complex sinusoid is a fundamental
signal processing problem which finds a lot of importance in fields
like radar signal processing.In radar signal processing the computational
capability not only needs to be accurate but also fast which involves
estimating the frequency every second.
The received signal is corrupted by white gaussian noise.
The best coarse frequency estimation of the signal is 
from the peak of the N point DFT of the received signal.
A N point DFT is typically calculated for a data length of N samples 
which gives a resolution of $\frac{2\pi}{N}$.
For real time spectral analysis a well known computationally efficient
method is the sliding DFT especially in the cases when a new DFT spectrum
is needed every few samples.The sliding DFT is computationally efficient than
the radix-2 FFT.
The sliding DFT performs a N point DFT within a sliding window of 
N samples.The window is then shifted by a sample for the next iteration
and a new N point DFT is calculated which utilises the old N point DFT
values.The single bin SDFT algorithm is implemented as an IIR filter,
to which a comb filter can be added so as to compute all N DFT spectral
components.In practical applications the algorithm can be initialised
with zero input and zero output.The output wont be valid until N samples
have been processed.
The sliding DFT though has a marginally stable transfer function because
all its poles lie on the $\emph{z}$-domain's unit circle.The filter coefficient
numerical rounding might force the poles outside of the unit circle which results
in instablity.A damping factor $\emph{r}$ may be used to force the pole into a 
radius of r inside the unit circle which guarantees stability.But the value of the
bins obtained from the guaranteed stable sliding DFT are different from that of the
traditional N point DFT.So the main problem lies in a fine estimation of frequency
from the guaranteed stable sliding DFT which differ from the traditional N point
DFT.
The fine frequency estimation generally follows the coarse frequency estimation which
is in turn done from the N point DFT.But the resolution of this process depends on the
spacing of points taken for the fine resolution.In [3]-[6] the fine resolution estimate
is done through a function on the DFT bins estimated from coarse resolution frequency
estimaton.In [7] a simple relation for DFT interpolation is used and in [8] a bias correction
is provided for [7].In this paper we derive Jacobsen's formula for the guaranteed stable
sliding DFT for which a bias correction term is also derived which includes bias correction
for the damping factor $\emph{r}$ as well.

\section{problem description}

A single complex sinusoid with white gaussian noise can be represented in the form
\emph{$$r[n]=Ae^{jwn}+w[n]$$}

where A and $\omega$ are unknown variables which represent the amplitude
and frequency of the complex sinusoid respectively
where $\omega$=$\frac{2\pi(k_p+\delta)}{N}$
and $k_p$ is the index of the peak of the sliding DFT.
$\delta$ is to be estimated from the three samples 
around the peak of the sliding DFT where $|\delta|$ $<$ 1/2 

The transfer function for N point sliding DFT filter can be represented as
\emph{$$H(z)=\frac{1-r^{N}z^{-N}}{1-re^{j2k\pi/N}z^{-1}}$$}
where $|r| <$ 1
Therefore a particular output bin of sliding DFT is written as

\emph{$$R[k]=\sum\limits_{n=0}^{N-1}x[n]r^{n}e^{-j2kn\pi/N}$$}
where $|r| <$ 1
In practice though a particular output bin can be found out using the following
recursive relation which basically serves the computational efficiency purpose
\emph{$$R_k[n]=R_k[n-1]re^{j2k\pi/N}-x(n-N)r^{N}+x(n)$$}
The sliding DFT bin where the peak occurs and its immediate neighbours can
be represented as follows:-
Let the indices for the peak be $k_p$ and that of its immediate neighbours
be $k_p-1$ and $k_p+1$ respectively.

\emph{$$R[k_p]=A\sum\limits_{n=0}^{N-1}r^{n}e^{j\frac{2\pi}{N}\delta n}+\hat{w}[k_p]$$}
              \emph{$$=Af(\delta)+\hat{w}[k_p]$$}        (1)
\emph{$$R[k_p-1]=A\sum\limits_{n=0}^{N-1}r^{n}e^{j\frac{2\pi}{N}(\delta+1) n}+\hat{w}[k_p-1]$$}
              \emph{$$=Af(\delta+1)+\hat{w}[k_p-1]$$}    (2)
\emph{$$R[k_p+1]=A\sum\limits_{n=0}^{N-1}r^{n}e^{j\frac{2\pi}{N}(\delta-1) n}+\hat{w}[k_p+1]$$}
              \emph{$$=Af(\delta-1)+\hat{w}[k_p+1]$$}    (3)      

where
$\hat{w}[k]$ is the DFT of w[n] which also is white and
\emph{$$f(\alpha)=\sum\limits_{n=0}^{N-1}r^{n}e^{j\frac{2\pi}{N}\alpha n}$$}.

Our aim is to estimate the value of $\delta$ from these three samples
$R[k_p]$,$R[k_p-1]$ and $R[k_p+1]$ so that $$\hat\omega =2\pi /N(k_p+\delta)$$ 
becomes the fine frquency estimate.
The two stage process consists of finding $k_p$ in the first stage and 
$\delta$ in the second stage.

\section{proposed estimator}

To determine $\delta$ from the set of three equations we consider the Geometric Progression
sums of each of the DFT bin and solve for $\delta$ using the approximation that the second and higher
powers of $\delta$ are negligible as compared to $\delta$.
We basically exploit the Geometric Progression sum in this case because of the added damping factor $r$
which in turn satisfies the relation $|r| <$ 1 and thus makes the common ratio in the Geometric Progression
to be less than one.

\emph{$$f(\alpha)=\sum\limits_{n=0}^{N-1}r^{n}e^{j\frac{2\pi}{N}\alpha n}$$}
                 \emph{$$=\frac{1-r^{N}}{1-re^{j\frac{2\pi \alpha}{N}}}$$}

where $|r|<$ 1
To estimate $\delta$
We evaluate the first difference and second difference of $f(\delta)$
For evaluting the first and second differences which are
$f(\delta+1)-f(\delta-1)$ and $f(\delta+1)-2f(\delta)+f(\delta-1)$ respectively (1) is used.
Using elementary trigonometry the first difference\emph{$$f(\delta+1)-f(\delta-1)$$}
\emph{$$=\frac{1-r^{N}}{1-re^{j\frac{2\pi (\delta+1)}{N}}}-\frac{1-r^{N}}{1-re^{j\frac{2\pi(\delta-1)}{N}}}$$}
which can be further simplified to
\emph{$$=\frac{2jr\sin (2\pi/N)e^{j2\pi \delta/N}(1-r^{N})}{1+r^{2}e^{j4\pi\delta/N}-2re^{j2\pi\delta/N}\cos(2\pi/N)}$$}
Using elementary trigonometry the second difference\emph{$$f(\delta+1)-2f(\delta)+f(\delta-1)$$}
\emph{$$=\frac{1-r^{N}}{1-re^{j\frac{2\pi (\delta+1)}{N}}}-\frac{2(1-r^{N})}{1-re^{j\frac{2\pi \delta}{N}}}+\frac{1-r^{N}}{1-re^{j\frac{2\pi (\delta-1)}{N}}}$$}
\emph{$$=\frac{(1-r^{N})(re^{j\frac{2\pi \delta}{N}}(e^{j\frac{\pi}{N}}- e^{-j\frac{\pi}{N}})
^{2}+r^{2}e^{j\frac{4\pi \delta}{N}}(e^{j\frac{\pi}{N}}- e^{-j\frac{\pi}{N}})^{2})}{(1+r^{2}e^{j4\pi\delta/N}-2re^{j2\pi\delta/N}\cos(2\pi/N))(1-re^{j2\pi \delta/N})}$$}

We get the following relation for the ratio of the two differences
\emph{$$\frac{f(\delta+1)-f(\delta-1)}{f(\delta+1)-2f(\delta)+f(\delta-1)}$$}
\emph{$$=\frac{-j\cot(\pi/N)(1-r^{2}-2jr\sin(2\pi \delta/N))}{1+r^{2}+2r\cos(2\pi \delta/N)}$$}
For large N and $\delta^{2}<<$1 we use the following approximations $\sin(2\pi \delta/N)=2\pi \delta/N$ and $\cos(2\pi \delta/N)=1$ the above relationship can be simplified to
\emph{$$=\frac{-j\cot(\pi/N)(1-r^{2}-4jr\pi \delta/N)}{(1+r)^2}$$}
\emph{$$=-\frac{4\pi r\delta \cot(\pi/N)}{N(1+r)^2}-\frac{j(1-r^{2})\cot(\pi/N)}{(1+r)^2}$$}
Simplifying the above relationship further we get
\emph{$$Real[\frac{f(\delta+1)-f(\delta-1)}{f(\delta+1)-2f(\delta)+f(\delta-1)}] $$}
\emph{$$=-\frac{4\pi r\delta \cot(\pi/N)}{N(1+r)^2}$$} (4)
Thus an estimate of $\delta$ can be produced by the substitution of
$f(\delta)\dashrightarrow R[k_p],f(\delta-1)\dashrightarrow R[k_p+1]$ and $f(\delta+1)\dashrightarrow R[k_p-1]$ 
(which were inturn derived in the relations (1),(2) and (3)) in (4);
\emph{$$\hat\delta=\frac{(1+r)^2}{4r}\frac{\tan(\pi/N)}{\pi/N}Real[\frac{R[k_p-1]-R[k_p+1]}{2R[k_p]-R[k_p-1]-R[k_p+1]}]$$}

\section{Numerical Comparisons}
In this section we present a numerical comparison between
the proposed estimator and the other estimators namely,
Candace's estimator and Jacobsen's estimator.The performance of 
the proposed estimator is compared across various damping factors which is
shown in Fig. 1.
The simulation is done in MATLAB.A sinusoidal signal is taken whose
frequency is varied from 30.1 MHz to 30.9 MHz in steps of 0.1 MHz.
128 samples of the signal are taken where  the sampling frequency is
128 MHz.A 128 point sliding DFT is taken for the proposed estimator while a 128 point
traditional DFT is taken for the other estimators.The noise taken is white gaussian noise.

\begin{figure}[h]
	\centering
		\includegraphics[width=90mm]{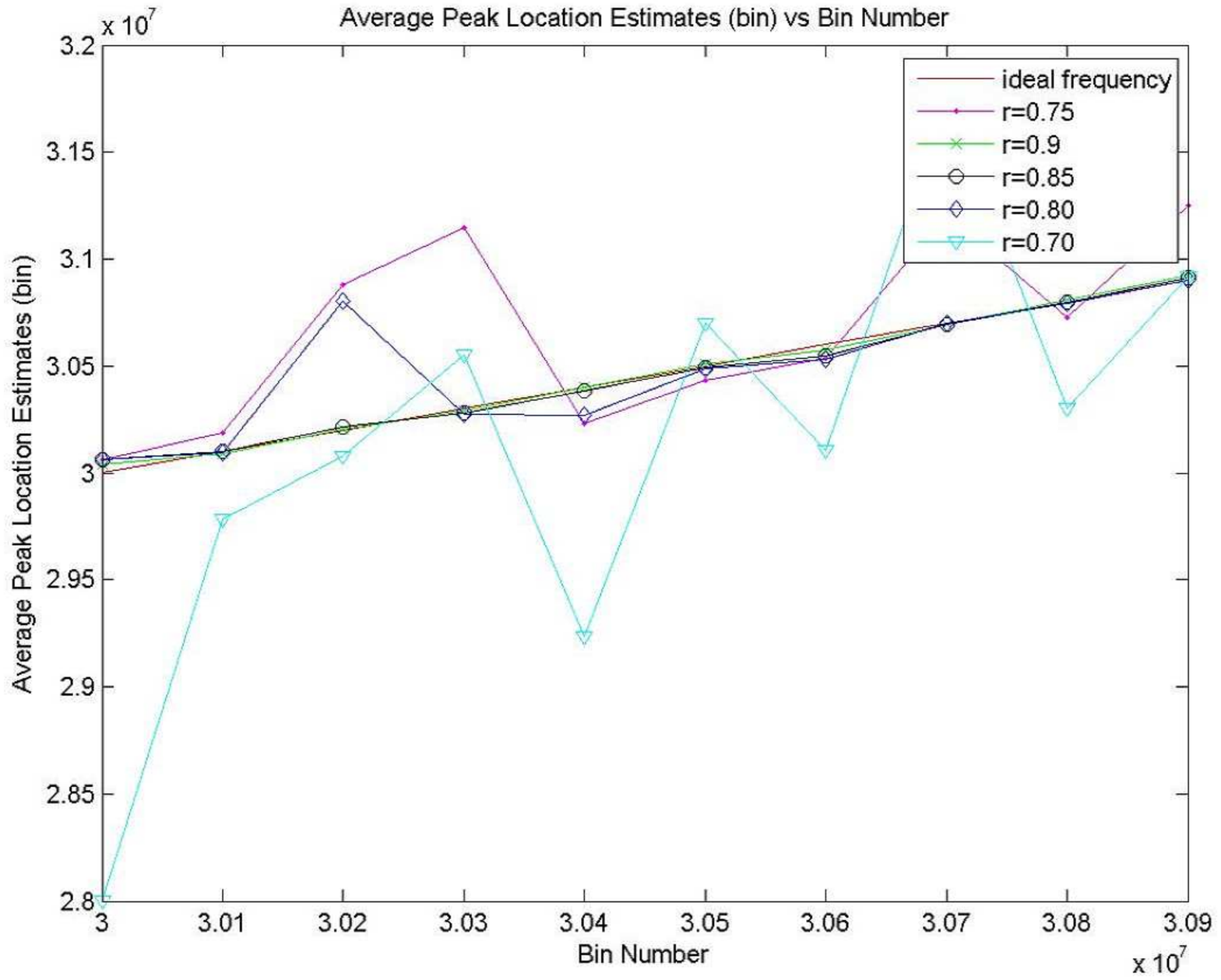}
	\caption{Performance of the proposed estimator with different damping factors}
\end{figure}

Fig. 2 shows the bias values of the proposed estimator and the other estimators with respect
to the ideal frequency in the absence of noise for $\emph{N}=$128.This shows that none of the
estimators are unbiased.For the proposed estimator the damping factor taken is $\emph{r}=$0.9
for the simulations.It can be clearly seen that the bias of the proposed estimator is less than
1000 Hz while those of the other estimators have higher bias values.
\begin{figure}[h]
	\centering
		\includegraphics[width=90mm]{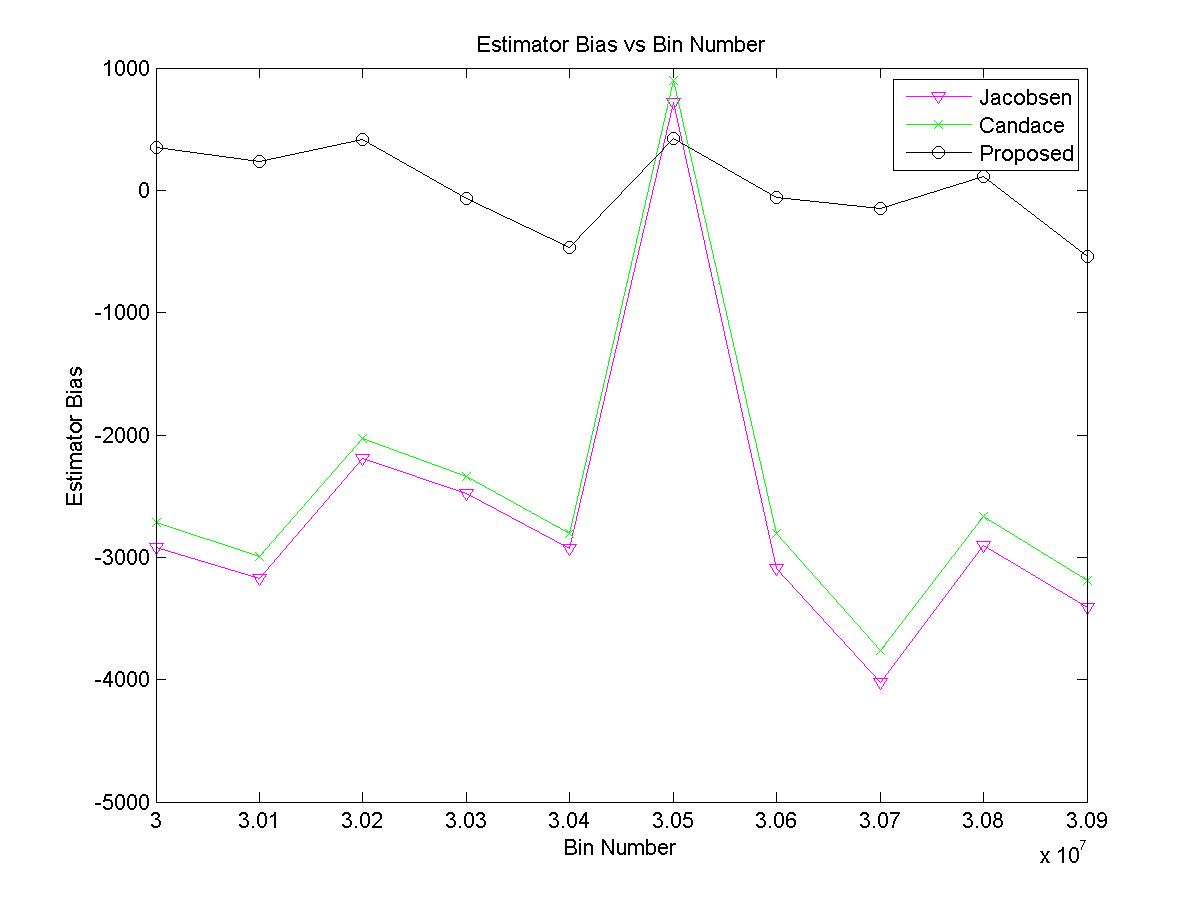}
	\caption{Variation of bias of different estimators in absence of noise}
\end{figure}

Fig. 3 shows the performance of the proposed estimator and the other estimators with respect to
ideal frequency is presence of noise for $\emph{N}=$128.
\begin{figure}[h]
	\centering
		\includegraphics[width=90mm]{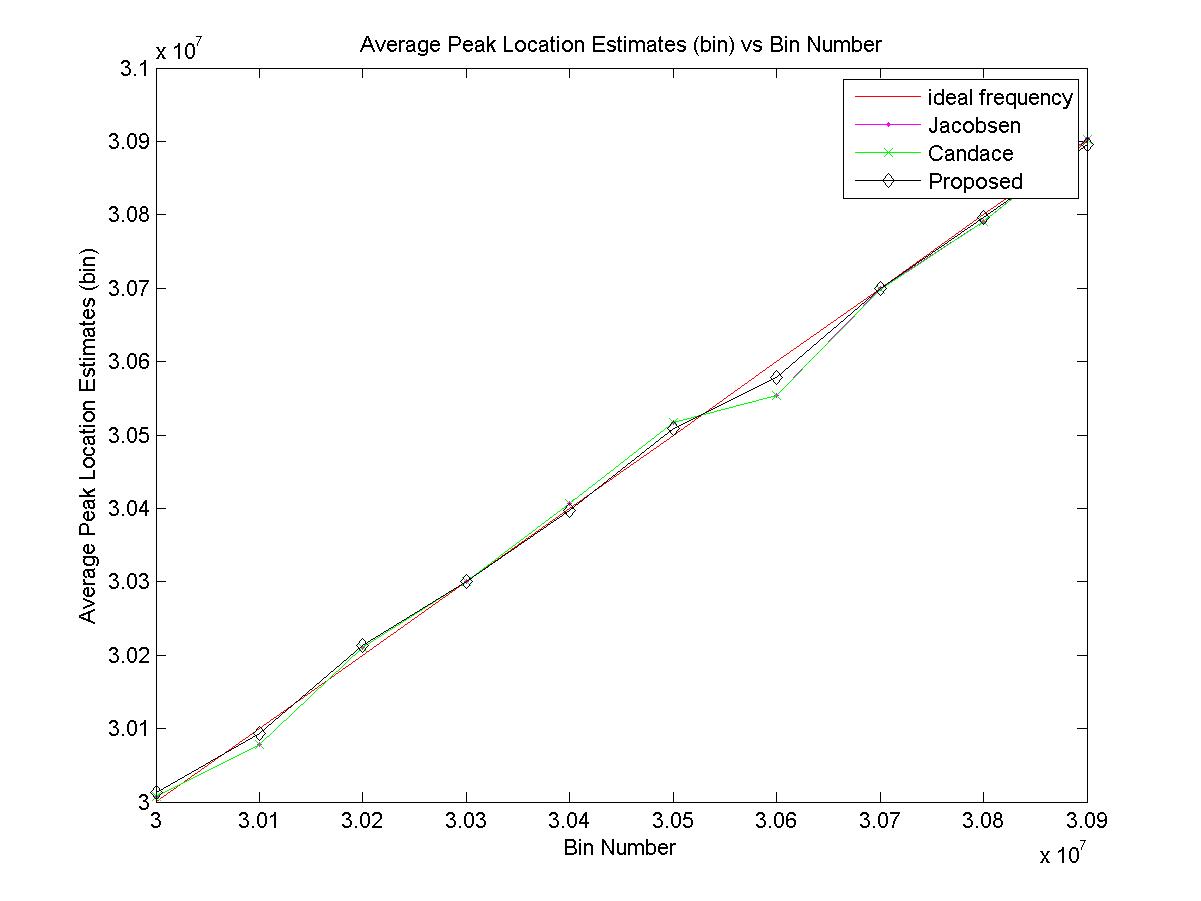}
	\caption{Performance of the estimators in presence of noise}
\end{figure}

Fig. 4 examines the bias of the estimators in presence of noise.It can be seen that the proposed
estimator has lower bias as compared to the other estimators.With higher SNR values the bias values
in presence of noise approach the bias values in absence of noise.
\begin{figure}[h]
	\centering
		\includegraphics[width=90mm]{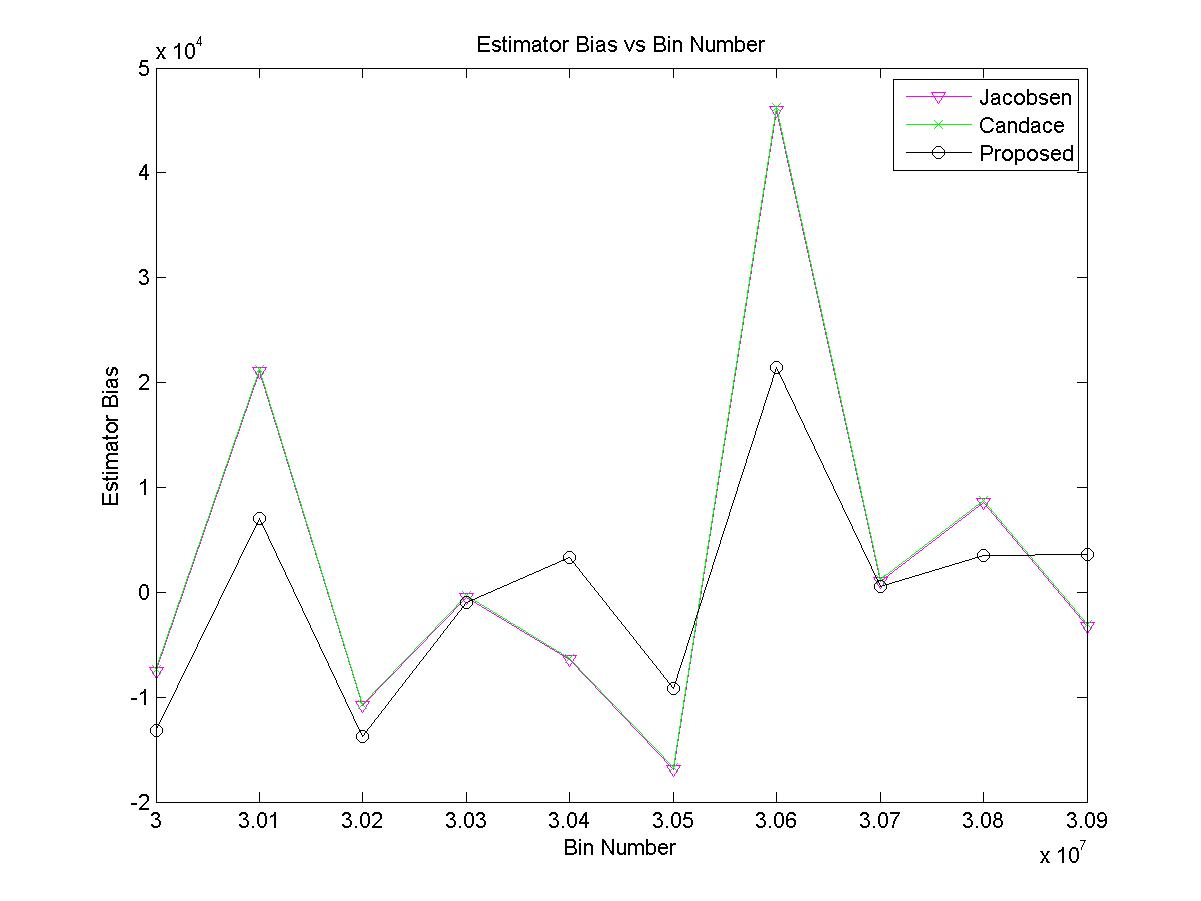}
	\caption{Variation of bias of different estimators in presence of noise}
\end{figure}

Fig. 5 examines the RMSE values of the estimators for large number of observations.

\begin{figure}[h]
	\centering
		\includegraphics[width=90mm]{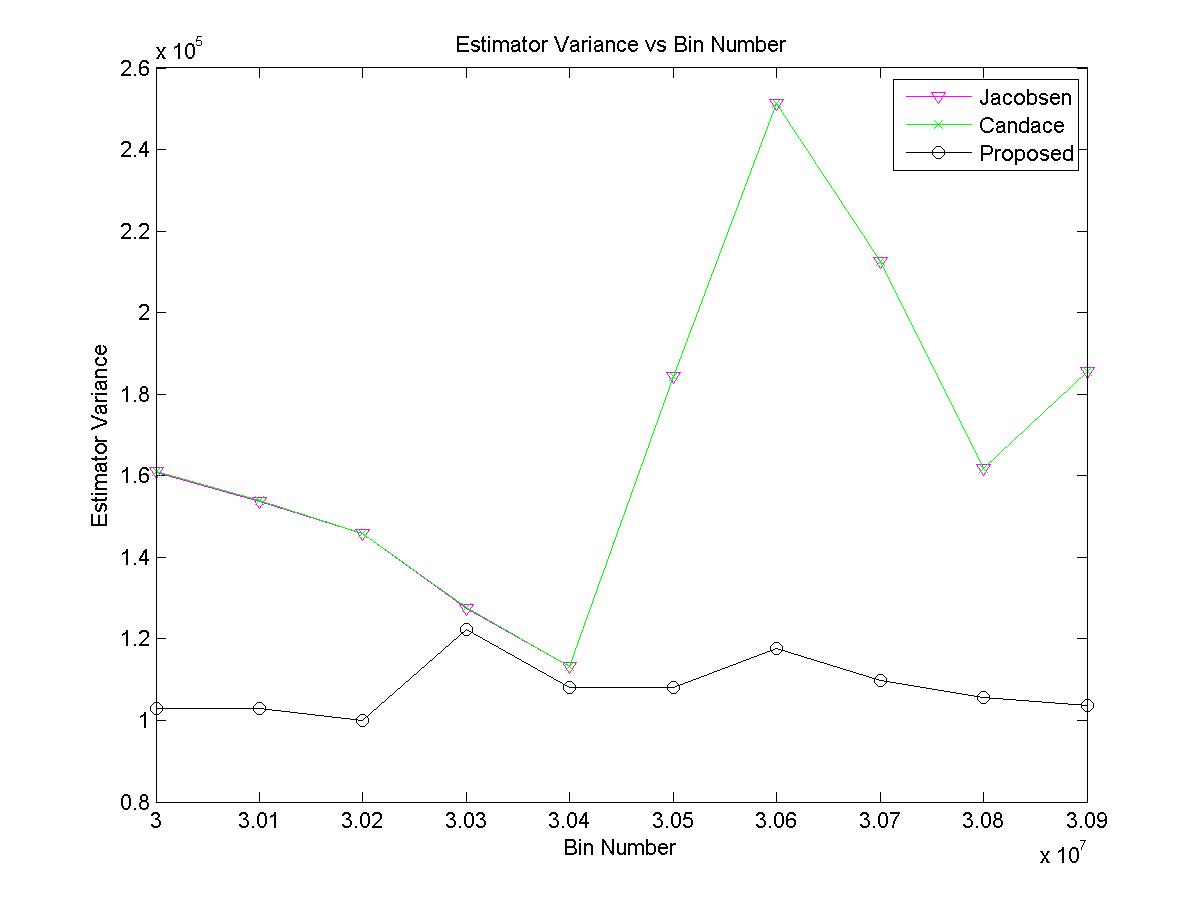}
	\caption{Variation of RMSE of different estimators in presence of noise}
\end{figure}

The SNR values taken for the above simulations range from 2 dB to 3dB.
At higher SNR values the estimator bias and the variance values decrease further and all the
three estimators behave nearly the same with the proposed estimator giving even lower bias and
variance values.
It should be noted that with the damping factor $\emph{r}$ approaching 1 with higher values of N
the proposed estimator have nearly the same performance.The value of the damping factor has been 
kept around 0.9 for guaranteed stability and better performance.It is intutively satisfying that
the bias correction factor in Candace's estimator is also a part of the derived expression for the
proposed estimator.The fewer number of operations involved for the proposed estimator compared to
that of other estimators and the lower bias and variance values makes the proposed estimator very useful
for radar signal processing. 
\section{Conclusions}
A new estimator is proposed which requires very few number
of operations per output sample.The estimator has a correction
term for bias.The good performance of the estimator is justified
in the paper.The proposed estimator has low bias and variance values 
which makes it a really valuable tool in the field of radar signal processing.
Simulation results showing superiority of the proposed estimator is provided.

The present work revolves around accurately estimating a single tone frequency or a multi tone frequencies
where the tones have large seperations.A potential future work is the extension of present work
to accurately estimating multi tone frequencies which are closely seperated without increasing the number
of computations.

\end{document}